# Nuclear catalysis mediated by localized anharmonic vibrations

Vladimir Dubinko

*National Science Center "Kharkov Institute of Physics and Technology", Kharkov 61108, Ukraine*
vdubinko@mail.ru

**Abstract**

In many-body nonlinear systems with sufficient anharmonicity, a special kind of lattice vibrations, namely, *Localized Anharmonic Vibrations* (LAVs) can be excited either thermally or by external triggering, in which the amplitude of atomic oscillations greatly exceeds that of harmonic oscillations (phonons) that determine the system temperature. Coherency and persistence of LAVs may have drastic effect on quantum tunneling due to *correlation effects* discovered by Schrödinger and Robertson in 1930. These effects have been applied to the tunneling problem by a number of authors, who demonstrated a giant increase of sub-barrier transparency during the increase of the *correlation coefficient* at a special high-frequency periodic action on quantum system. Recently, it has been proposed that *discrete breathers* (a sub-class of LAVs arising in *periodic systems*) present the most natural and efficient way to produce correlation effects in *regular crystals* due to time-periodic modulation of the potential well (or the Coulomb barrier) width and hence to act as breather 'nano-colliders' catalyzing low energy nuclear reactions (LENR) in solids. It has been shown that the tunneling probability for the D-D fusion under electrolysis of heavy water increases enormously with increasing number of oscillations resulting in the fusion rates comparable with those observed experimentally. In the present paper, we discuss possible ways of *engineering the nuclear-active environment (NAE)* and *catalyzing LENR in NAE* based on the LAV concept. We propose some practical ways of catalyzing LENR that are based on a special electro-magnetic treatment or electron irradiation, which trigger LAVs in crystals and clusters.

*Keywords:* localized anharmonic vibrations, discrete breathers, correlation effects, low energy nuclear reactions, nuclear active sites.

_______________________________________________________________________________________

## Content

## 1. Introduction

Catalysis is at the heart of almost every chemical or nuclear transformation process, and a detailed understanding of the active species and their related reaction mechanism is of great interest [1-3]. There is no single theory of catalysis, but only a series of principles to interpret the underlying processes. An important parameter of the reaction kinetics is the *activation energy*, i.e. the energy required to overcome the reaction barrier. The lower is the activation energy, the faster the reaction rate, and so a catalyst may be thought to reduce somehow the activation energy. Dubinko et al have shown [4, 5] that in a crystalline matrix, the activation energy may be reduced at some sites due to a special class of *localized anharmonic vibrations* (LAV) of atoms, known also as *discrete breathers* [6] or *intrinsic localized modes* [7] arising in regular crystals. Discrete breathers (DB) can be excited thermally or by external driving, resulting in a drastic acceleration of *chemical reaction* rates in their vicinity. What is more, recently, discrete breathers have been proposed as **catalysts of nuclear D-D fusion** in palladium deuteride under heavy water electrolysis [8]. This new concept does not require medication of conventional nuclear physics lows, in a marked contrast to the most of LENR models that attempted introducing various types of transient quasi-particles and structures such as Hydrino, Hydron, Hydrex etc. that were expected to lower the Coulomb barrier.

One of the most important practical recommendations of the new concept is to look for the environment, which can be enriched with sites of *zero or small threshold energies* for the excitation of DBs. Such sites are expected to become the *nuclear active cites*, according to the model [8]. In this context, a striking *site selectiveness* of LAV formation in disordered structures [5] allows one to suggest that their concentration in quasicrystals may be very high as compared to regular crystals where DBs (a sub-class of LAV) arise homogeneously, and their activation energy is relatively high. Direct experimental observations [9] have shown that in the decagonal quasicrystal Al$_{72}$Ni$_{20}$Co$_8$, *mean-square thermal vibration amplitude* of the atoms at special sites greatly exceeds the mean value, and the difference increases with temperature. This might be the first experimental observation of LAV, which has shown that they are arranged in just a few nm from each other, which means that their average concentration was about $10^{20}$ per cubic cm that is orders of magnitude higher than one could expect to find in periodic crystals [4, 5, 8]. So in this case, one deals with a kind of *'organized disorder'* that stimulates formation of LAV, which may explain a strong catalytic activity of some quasicrystals and open new ways towards *engineering of NAE* based on computer modeling of LAVs in periodic and aperiodic crystals and nano clusters. The main goal of the present paper is to develop this concept to the level of quantitative comparison with some of the LENR experiments and to suggest some practical ways of catalyzing LENR.

The paper is organized as follows. In the next section, we discuss shortly the problem of tunneling through the Coulomb potential barrier taking into account the *correlation effects*, as proposed by Dodonov et al [10] and more recently by Vysotskii et al [11]. A giant increase of sub-barrier transparency (up to hundreds orders of magnitude) during the increase of the correlation coefficient at special periodic action on quantum system will be demonstrated.

In section 3, we will show that such an action can be most naturally realized due to time-periodic modulation of the width of potential wells for atoms oscillating in the vicinity of a *DB* or, more generally, LAV, and find the critical parameters of LAVs required to form *coherent correlated states* (CCS) via 'parametric resonance' conditions.

In section 4, based on the LAV concept, an alternative to heating (currently used for triggering LENR) is proposed that is based on a special electro-magnetic treatment or electron irradiation, which can trigger LAVS in crystals and clusters and thus catalyze LENR. The ways of experimental verification of the proposed concept are discussed in section 5. The summary and outlook is given in section 6.

## 2. Tunneling enhancement due to formation of CCS in non-stationary potential well

*2.1 Heisenberg Uncertainty Relation, Gamow TC and screening potential effects*

The problem of tunneling through the Coulomb potential barrier during the interaction of charged particles is the key to modern nuclear physics, especially in connection with low energy nuclear reactions (LENR) observed in solids [12-14].

The tunneling (a.k.a. transmission) coefficient (TC) first derived by Gamow (1928) for a pure Coulomb barrier is the Gamow factor, given by

$$G \approx \exp\left\{-\frac{2}{\hbar}\int_{r_1}^{r_2} dr \sqrt{2\mu(V(r)-E)}\right\} \qquad (1)$$

where $2\pi\hbar$ is the Planck constant, $E$ is the nucleus CM energy, $\mu$ is the reduced mass, $r_1$, $r_2$ are the two classical turning points for the potential barrier, which for the D-D reaction are given simply by $\mu = m_D/2$, $V(r) = e^2/r$. For two D's at room temperature with thermal energies of $E \sim 0.025$ eV, one has $G \sim 10^{-2760}$, which explains a pessimism about LENR and shows the need for some *special conditions* arising in solids under typical LENR conditions (D$_2$O electrolysis [12, 13] or E-cat [14], etc.), which help to overcome the Coulomb potential.

Corrections to the cross section of the fusion due to the screening effect of atomic electrons result in the so-called "*screening potential*", which acts as an additional energy of collision at the center of mass [15]. The screening potential was measured by the yields of protons or neutrons emitted in the D(d, p)T or D(d, n)$^3$He reactions induced by bombardment of D-implanted solid targets with deuterons accelerated to kinetic energies of several keV, equivalent to heating up to ~$10^7$ K [16]. However, even the maximum screening potentials found in Pt (675 eV), PdO (600 eV) and Pd (310 eV) are far too weak to explain LENR observed at temperatures, which are bellow melting point of solids (E-cat) or boiling point of liquids (electrolysis). Besides, the absence of significant radiation under typical LENR conditions indicates that **other reactions** should take place, based on interactions between 'slow' particles, which may be qualitatively different from the interactions between accelerated ones.

The most promising and universal mechanism of the stimulation of nuclear reactions running at a low energy is connected with the formation of *coherent correlated states* of interacting particle, which ensures the large probability of the nuclear reactions under conditions, where the ordinary tunneling probability is negligible. These states minimize a more general uncertainty relation (UR) than Heisenberg UR usually considered in quantum mechanics, namely, Schrödinger-Robertson UR [17, 18], which takes into account *correlations between coordinate and momentum operators*.

*2.2 Schrödinger-Robertson Uncertainty Relation and TC*

The tunneling effect for nuclear particles is closely related to the uncertainty relation (UR), which determines, in fact, the limits of the applicability of the classical and quantum descriptions of the same object. It appears that the well-known and widely used Heisenberg UR is a special case of a more general inequality, discovered independently by Schrödinger [17] and Robertson [18], which can be rewritten in the following form [10, 19]

$$\sigma_x \sigma_p \geq \frac{\hbar^2}{4(1-r^2)}, \quad r = \frac{\sigma_{xp}}{\sqrt{\sigma_x \sigma_p}}, \quad \sigma_{xp} = \langle \hat{x}\hat{p} + \hat{p}\hat{x} \rangle / 2 - \langle x \rangle \langle p \rangle \tag{2}$$

$$\sigma_x = \langle (x - \langle x \rangle)^2 \rangle, \quad \sigma_p = \langle (p - \langle p \rangle)^2 \rangle, \tag{3}$$

where, $r$ is the *correlation coefficient* between the coordinate, $x$, and momentum, $p$, operators. At $r = 0$ (non-correlated state) eq. (2) is reduced to the Heisenberg UR, while in a general case, a nonzero $r$ in the UR can be taken into account by the formal substitution

$$\hbar \to \hbar_{ef} = \frac{\hbar}{\sqrt{1-r^2}}, \tag{4}$$

which leads to the formal shift of the border between the classical and quantum descriptions of the same object in the transition from non-correlated to correlated state [11].

Then a natural question arises: can nonzero correlations lead to real physical effects? **The answer is yes**, and the most impressive consequence is a dramatic increase of the tunneling probability, if a true Planck constant in eq. (1) can be replaced by the effective parameter (4). This substitution was indeed justified for a very low tunneling probability in the initially uncorrelated state $G_{r=0} \ll 1$ that corresponds to the condition $E \ll V_{\max}$ [11]:

$$G_{r \neq 0} \approx \exp\left\{ -\frac{2}{\hbar_{ef}} \int_{r_1}^{r_2} dr \sqrt{2\mu (V(r) - E)} \right\} = (G_{r=0})^{\sqrt{1-r^2}}, \tag{5}$$

which is close (within an order of magnitude) to the result of the *exact calculation* of the potential barrier transparency using rigorous quantum-mechanical methods [11]. From eq. (5), it follows that when a strongly correlated state with $|r| \to 1$ is formed, the product of the variances of the particle coordinate and momentum increases indefinitely, and the barrier becomes practically 'transparent': $G_{|r| \to 1} \to 1$ even if $E \ll V_{\max}$.

Although the substitution $\hbar \to \hbar_{ef}$ (4) is not quite correct quantitatively, it clearly demonstrates the high efficiency of the correlations in applied tunneling-related problems in the case of a high potential barrier and a low particle energy $E \ll V_{\max}$, which is typical for LENR.

The physical reason for the huge increase in barrier transparency for a particle in a CCS is the *co-phasing* of all fluctuations of the momentum for various eigenstates forming the superpositional correlated state, which leads to great dispersion of momentum and large fluctuations of positions of a particle in the potential well and *under the potential barrier*.

A CCS can be formed in various quantum systems. The most effective way to form such state is to place a particle in a *non-stationary potential well*. Exact solutions to the non-stationary Schrödinger equation for the oscillating particle wave function $\psi(x,t)$ have been found by Vysotskii et al [11], which will be considered in the following section.

*2.3 Formation of CCS under time-periodic action on a particle in the parabolic potential*

Consider a harmonic oscillator in a non-stationary potential well with eigenfrequency $\omega(t)$ changing with time, $t$, the Hamiltonian of which is given by

$$\hat{H} = \frac{\hbar}{2M} \frac{d}{dx} + \frac{M\omega^2(t)\hat{x}^2}{2} \tag{6}$$

The solution of the Schrödinger equation for this Hamiltonian by Dodonov and Man'ko [10, 19] gives the wave function of the correlated state having the time-periodic oscillating dispersion of the particle coordinate $\sigma_x(t)$:

$$\Psi_{corr}(x,t,r) = \frac{1}{\sqrt[4]{2\pi\sigma_x(t)}} \exp\left[ -\frac{x^2}{4\sigma_x(t)} \left( 1 - \frac{ir(t)}{\sqrt{1-r^2(t)}} \right) \right], \quad \sigma_x(t) = \frac{x_0^2}{2\sqrt{1-r^2(t)}} \tag{7}$$

where $x_0 = \sqrt{\hbar/M\omega_0}$ is the amplitude of zero-point oscillations, which shows the scaling of the quantum problem. With increasing $|r| \to 1$, the dispersion increases indefinitely, which means that a probability of the particle localization in the sub-barrier region increases accordingly (Fig. 1). For example, $x_0 \approx 0.56\,\text{Å}$ for a hydrogen nucleus, and the probability of its tunneling to a distance comparable with the crystal lattice spacing (~ angstroms) drastically increases with increasing $|r|$.

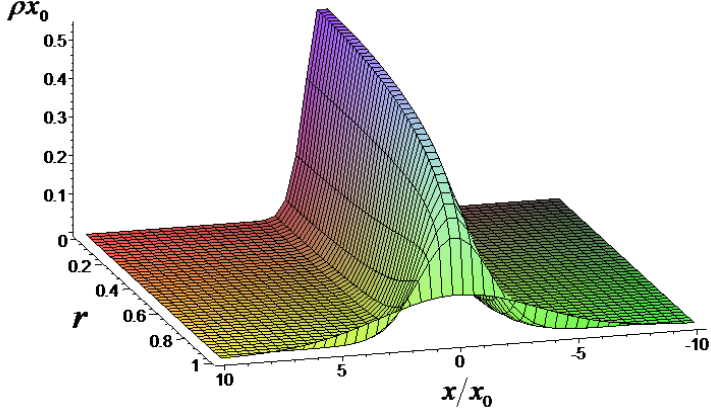

**Fig.1.** Product of the probability density $\rho = |\psi(x,r)|^2$ and the characteristic width of the harmonic potential well, which illustrates the particle localization in the well and in the sub-barrier region at different r values [20].

A model system considered by Vysotskii et al [11] for evaluation of the correlation coefficient is a particle with the mass $M$, coordinate $x(t)$ and momentum $p(t)$ in a non-stationary parabolic potential well (i.e. non-stationary harmonic oscillator),

$$V(x,t) = M(x(t))^2 (\omega(t))^2 /2, \qquad (8)$$

for which a change of the eigenfrequency $\omega(t)$ was shown to result in an increase of the absolute value of $|r(t)|$. Several scenarios of time evolution $\omega(t)$ have been investigated [11], including its monotonic decrease or periodic modulation. The latter regime can be provided, e.g. at a constant well depth $V_{\max}$ if the potential well width $L(t)$ changes periodically resulting in a time-periodic modulation of the eigenfrequency as follows:

$$L(t) = L_0 (1 + g_\Omega \cos \Omega t), \, L_0 = \sqrt{8V_{\max}/M\omega_0^2}, \qquad (9)$$

where $L_0$ and $\omega_0$ are the initial parameters of the well before the action of correlated forces, $g_\Omega$ and $\Omega$ are the modulation amplitude and frequency, respectively.

From a detailed analysis [11] it follows that the process of formation of strongly correlated coherent state with $|r|_{\max} \to 1$ in response to the action of limited periodic modulation is possible only at any of two conditions: (i) $\Omega = \omega_0$ (resonant formation) or (ii) $\Omega$ is close to $2\omega_0$ (parametric formation): $|\Omega - 2\omega_0| \leq g_\Omega \omega_0$.

Fig. 2 shows evolution of the correlation coefficient in time under the action of the harmonic perturbation with frequencies $\Omega = \omega_0$ and $\Omega = 2\omega_0$ at $g_\Omega = 0.1$. Correlation coefficient oscillates with time but its amplitude $|r|_{\max}$ increases monotonously with the number of modulation cycles, $n = \omega_0 t / 2\pi$, resulting in a giant increase of the tunneling coefficient, as demonstrated in Fig. 3, which shows the TC evaluated by eq. (10) that takes into account both the electron screening and the correlation effects [8]:

$$G^*(L,r) = \exp\left\{-\frac{2\pi e^2}{\hbar_{ef}(r)} \sqrt{\frac{\mu}{2(E + e^2/L)}}\right\}, \qquad (10)$$

where $L$ is the minimum equilibrium spacing between D atoms determined by electron screening, $E$ is their kinetic energy (~eV/40 at room temperature) << screening energy~ $e^2/L$. One can see that the difference in electron screening and the corresponding initial D-D distances in a $D_2$ molecule ($L_0 = 0.74$ Å) and in the PdD crystal ($L_0 = 2.9$ Å) leads to a huge tunneling difference in the initial (uncorrelated) state, in which TC is negligible in both cases. However, with increasing number of modulation cycles, $\hbar_{ef}(r)$ increases according to Fig. 2 resulting in a giant increase of TC up to ~1 in several hundreds of cycles for $\Omega = \omega_0$ and in several dozens of cycles for parametric formation $\Omega \approx 2\omega_0$, which is more realistic to attain since it does not require exact coincidence of the frequencies [11].

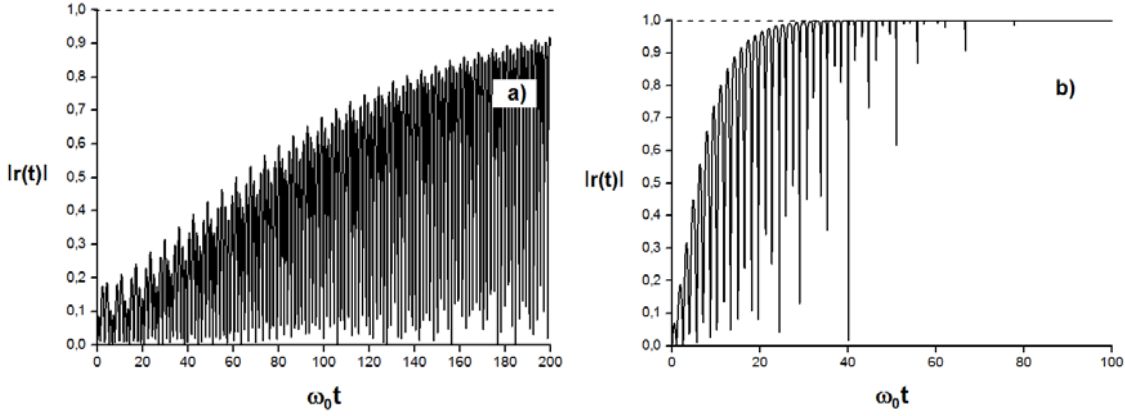

**Fig.2.** Correlation coefficient vs. time of action of the harmonic modulation of the well width given by eq. (10) at $g_\Omega = 0.1$ and two modulation frequencies: **a)** $\Omega = \omega_0$ - resonant frequency, and **b)** $\Omega = 2\omega_0$ - parametric resonance frequency [11].

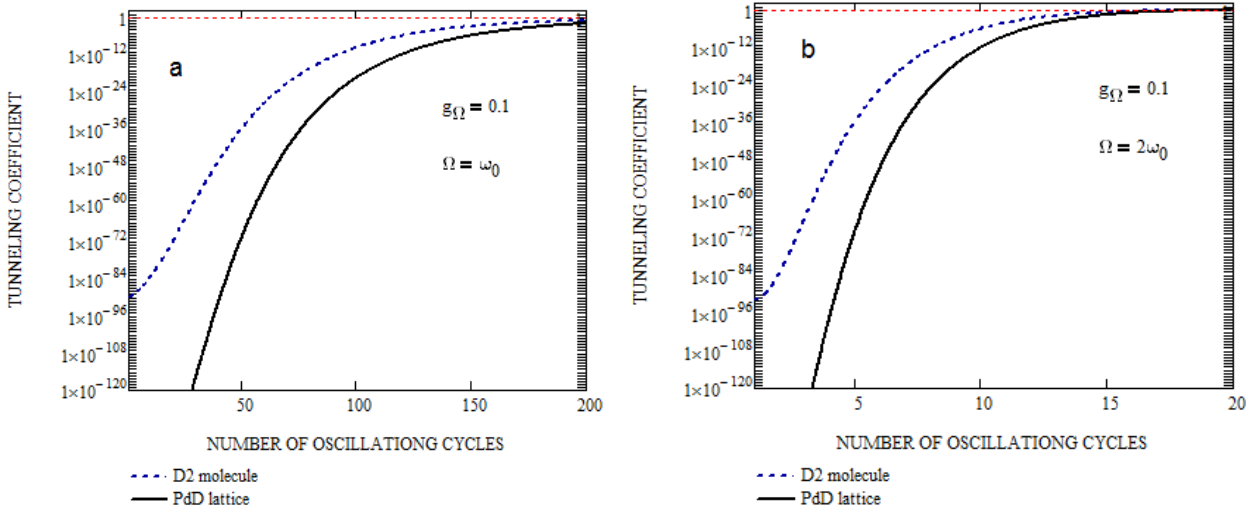

**Fig.3.** Tunneling coefficient increase with increasing number of the well modulation cycles, $n = \omega_0 t / 2\pi$, evaluated by eq. (10) for $\Omega = \omega_0$ (a); $\Omega \approx 2\omega_0$ (b), $g_\Omega = 0.1$ for two D-D equilibrium spacings: in a $D_2$ molecule ($L_0 = 0.74$ Å) and in the PdD crystal ($L_0 = a_{PdD}\sqrt{2}/2 \approx 2.9$ Å). $a_{PdD} = 4.052$ Å is the PdD lattice constant at 295 K [8].

The most important and nontrivial practical question now is how to realize such a periodic action at atomic scale? Modulation of the frequency of the optical phonon modes via excitation of the surface electron plasmons by a terahertz laser suggested in [11] as a driving force for the CCS formation is very questionable [8], and it does not explain LENR observed in the absence of the laser driving. In the next section, we will consider a mechanism based on the large-amplitude time-periodic oscillations of atoms, which is an inherent LAV property.

### 3. LAV-induced time-periodic action on the potential barrier

*3.1 Discrete breathers in bulk periodic crystals*

The first attempts to develop a LENR mechanism in metal hydrides/deuterides (e.g. PdD or NiH) [8] were based on their crystal *periodic* structure, characteristic for *bulk* specimens or large particles. At ambient conditions, Pd hydrides/deuterides crystallize in FCC structure with the space group of the Rock-salt [21], while Pd hydrides/deuterides crystallize in various structures corresponding to NiH, NiH2 and NiH3 [22]. Molecular dynamic (MD) simulations have revealed that diatomic crystals with Morse interatomic interactions typically demonstrate *soft type* of anharmonicity [23], which means that DB's frequency *decreases* with increasing amplitude, and one can expect to find so-called gap DBs with frequencies within the phonon gap of the crystal. The large mass difference between H or D and the metal atoms is expected to provide a wide gap in phonon spectrum, in which DBs can be excited either by *thermal fluctuations* at elevated temperatures (as demonstrated for the different weight ratios and temperatures [24]), or by some *external driving* providing a sufficiently large initial displacement of a light atom from its equilibrium position (Fig. 4).

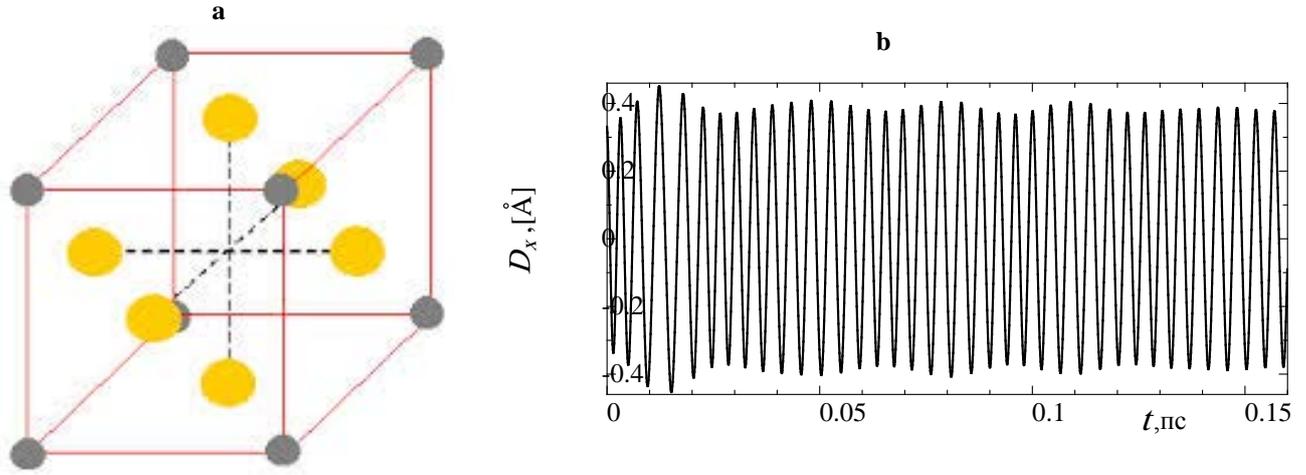

**Fig.4. (a)** $A_3B$ compound based on FCC lattice with Morse interatomic potentials. Grey atoms are 50 times lighter than yellow ones. **(b)** DB initiated at zero Kelvin by the initial displacement $D_{x0} \sim 0.3$ Å of a light atom from its equilibrium position. It is localized on a **single light atom** vibrating along <100> direction with the frequency of 227 THz, which lies **inside the phonon gap**. Shown is the x-displacement of the light atom as the function of time. DB has a large amplitude of ~0.4 Å that should be compared to the lattice parameter of 1.35 Å [25].

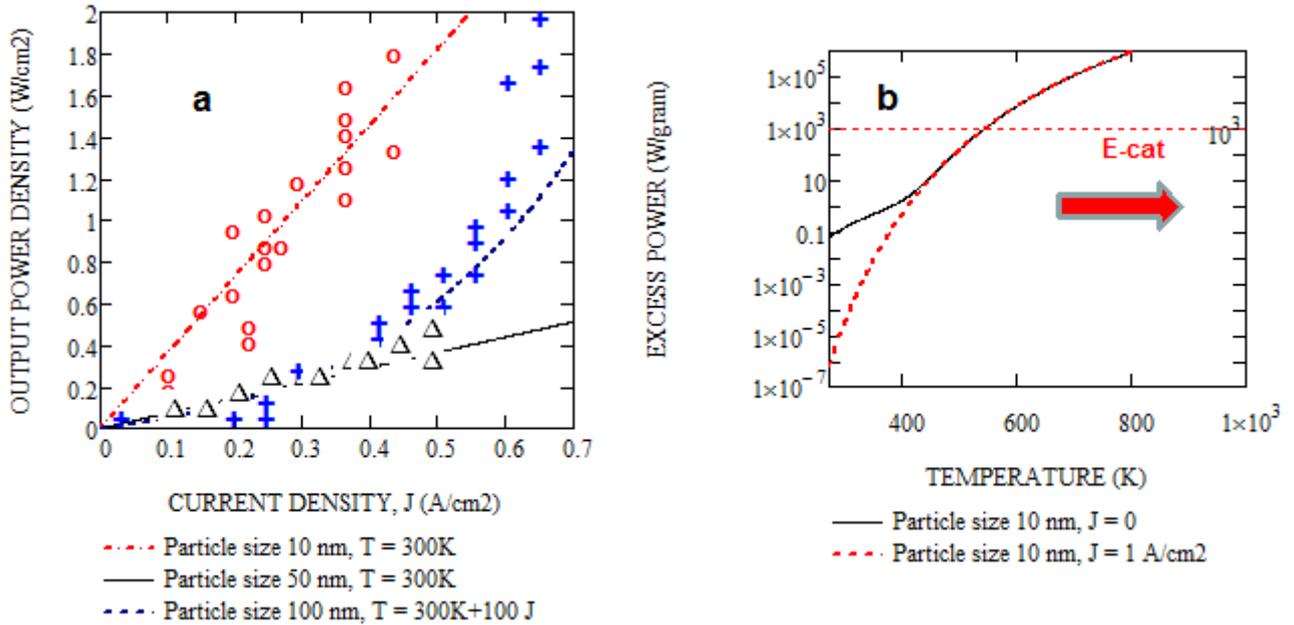

**Fig.5**. **(a)** LENR output power density according to the model [8] vs. experimental data [13] as a function of electric current density at constant *T* and at *T* increasing with *J* as *T*=300K+100*J*. **(b)** LENR output power density according to the model [8] vs. temperature (dashed line shows the E-cat power generation level [14]).

Thus, the D-D fusion rate in *bulk* PdD should be determined by the generation rate of DBs having amplitudes near the critical value ~ 0.3 Å, which has been evaluated in [8] for typical LENR conditions, resulting in a qualitative and quantitative agreement with experimental data on fusion rate under $D_2O$ electrolysis at the fitted set of well-defined material parameters (Fig. 5a). In particular, the model explains the crucial role of electric current under electrolysis that produces the flux of energetic deuterons (in the eV range) through the cathode surface, which is a driving force for the DB generation. In the absence of electric current, the LENR rate is predicted to be lower by 5 orders of magnitude (Fig 5b).

With increasing temperature, the concentration of DB increases resulting in the increased power generation, in a *qualitative agreement* with E-cat experiments [14] (Fig. 5b). However, the model predicts much more powerful LENR output above 600 C than what is observed in E-cat type installations [14]. This discrepancy appears because the model [8] did not take into account effect of *temperature fluctuations* on the tunneling efficiency of DBs. As has been shown by Vysotskii et al [26], fluctuations interfere with formation of CCS, which means that the ***temperature effect is two-fold and controversial***. On the one side, heating helps to excite DBs but on the other hand, it decreases their tunneling efficiency. Therefore, the best way to facilitate LENR, according to the present concept, is to look for environment, which favors formation of DBs (or, more generally - LAVs) at lower temperatures than those explored in typical E-cat installations. In the following section, we consider LAVs in metal hydride/deuteride nanocrystals.

*3.2 LAVs in nanocrystals and quasicrystals*

The fact that the energy localization manifested by LAV does not require long-range order was first realized as early as in 1969 by Alexander Ovchinnikov who discovered that localized long-lived molecular vibrational states may exist already in simple molecular crystals ($H_2$, $O_2$, $N_2$, NO, CO) [27]. He realized also that stabilization of such excitations was connected with the *anharmonicity* of the intramolecular vibrations. Two coupled anharmonic oscillators described by a simple set of dynamic equations demonstrate this idea:

$$\ddot{x}_1 + \omega_0^2 x_1 + \varepsilon \lambda x_1^3 = \varepsilon \beta x_2$$
$$\ddot{x}_2 + \omega_0^2 x_2 + \varepsilon \lambda x_2^3 = \varepsilon \beta x_1 \quad , \quad (11)$$

where $x_1$ and $x_2$ are the coordinates of the first and second oscillator, $\omega_0$ are their zero-point vibrational frequencies, $\varepsilon$ is a small parameter, and $\lambda$ and $\beta$ are parameters characterizing the *anharmonicity* and the *coupling* force of the two oscillators, respectively. If one oscillator is displaced from the equilibrium and start oscillating with an initial amplitude, *A*, then the time needed for its energy to transfer to another oscillator is given by the integral:

$$T = \frac{\omega_0}{\varepsilon \beta} \int_0^{\pi/2} \frac{d\varphi}{\sqrt{1 - \left(A^2 \gamma / 4\right)^2 \sin^2 \varphi}}, \qquad \gamma = \frac{3\lambda}{\beta}, \quad (12)$$

from which it follows that the full exchange of energy between the two oscillators is possible only at sufficiently small initial amplitude: $A^2 \gamma / 4 < 1$. In the opposite case, $A^2 \gamma / 4 > 1$, the energy of the first oscillator will *always be larger* than that of the second one. And for sufficiently large initial amplitude, $A \gg \sqrt{4/\gamma}$, there will be practically no sharing of energy, which will be localized exclusively on the first oscillator. This remarkable and seemingly *contra-intuitive* result is illustrated in Fig. 6 that shows the phase diagram of two coupled anharmonic oscillators.

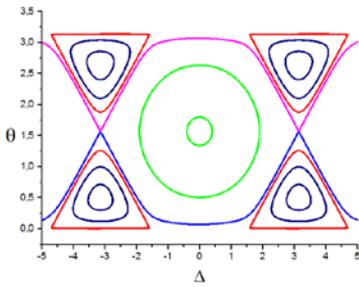

**Fig.6.** Phase diagram of two coupled anharmonic oscillators, according to the Ovchinnikov model [27] (see text). Separate regions in the diagram correspond to phase trajectories of the oscillators that *cannot share energy* [20].

Thus, Ovchinnikov has proposed the idea of LAV for molecular crystals, which was developed further for any nonlinear systems possessing *translational symmetry*; in the latter case, LAVs have been named *discrete breathers* (DBs) [6] [1] or *intrinsic localized modes* (ILMs) [7]. Now, we are coming back to the idea of LAV arising at 'active sites' in defected crystals, quasicrystals and nanoclusters. As noted by Storms [13], 'Cracks and small particles are the Yin and Yang of the cold fusion environment'. A physical reason behind this phenomenology is that in topologically disordered systems, sites are not equivalent and band-edge phonon modes are intrinsically localized in space. Hence, different families of LAV may exist, localized at different sites and approaching different edge normal modes for vanishing amplitudes [5]. Thus, in contrast to perfect crystals, which produce DB homogeneously, there is a striking *site selectiveness* of energy localization in the presence of spatial disorder, which has been demonstrated by means of atomistic simulations in biopolymers [5], metal nanoparticles [29] and, *experimentally*, in a decagonal *quasicrystal* Al$_{72}$Ni$_{20}$Co$_{8}$ [9]. In the latter case, the authors measured a so-called Debye–Waller factor defined by the mean-square thermal vibration amplitude of the atoms, and demonstrated that the anharmonic contribution to Debye–Waller factor increased with temperature much stronger than the harmonic (phonon) one. This was the first <u>direct observation</u> of a 'local thermal vibration anomaly' i.e. LAV, in our terms. The experimentally measured separation between LAVs was about 2 nm, which meant that their mean concentration was about $10^{20}$ per cm$^3$ that is many orders of magnitude higher than one could expect to find in periodic crystals [8].

---

[1] Note that the term *discrete breathers* was coined nearly 20 years ago [6] in western nonlinear community, contrary to the recent comments [28] assigning the term to Ukrainian literature, while in the present paper we introduce a new and more general term 'LAV'.

The crystal shape of the nanoparticles (cuboctahedral or icosahedral) is known to affect their catalytic strength [30], and the possibility to control the shape of the nanoparticles using the amount of hydrogen gas has been demonstrated both experimentally by Pundt et al [31], and by means of atomistic simulations by Calvo et al [32]. They demonstrated that above room temperature the *icosahedral phase* should remain stable due to its higher entropy with respect to cuboctahedron (Fig. 7). And icosahedral structure is one of the forms quasicrystals take, therefore one is tempted to explore further the *link between nanoclusters and quasicrystals*.

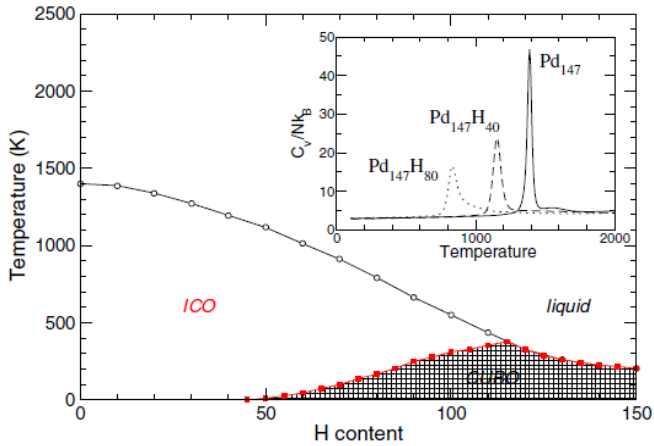

**Fig. 7**. Schematic structural diagram of the Pd$_{147}$H$_x$ cluster in the icosahedral, cuboctahedral and liquid phases, after [32]. Inset: heat capacities of three clusters, in units of k$_B$ per atom, versus canonical temperature. Icosahedral phase is predicted to be more stable above room temperature.

Fig. 8 shows the structure of Pd$_{147}$H$_{138}$ cluster containing 147 Pd and 138 H atoms having minimum free energy configuration, replicated using the method and parameters by Calvo et al [32]. In particular, Fig. 8(**b**) reveals the presence of H-H-H chains aligned along the I-axis of the cluster. This *ab initio* simulation points out at the possibility of excitation of LAVs in these chains, with a central atom performing large-amplitude anharmonic oscillations and its neighbors oscillating in quasi-harmonic regime [34], which is similar to that considered in [8] for regular diatomic lattice of NaCl type. Such oscillations have been shown to stimulate formation of CCSs [8] and to facilitate LENR. In the following section, we will consider the ways of excitation of LAVs based on the present concept.

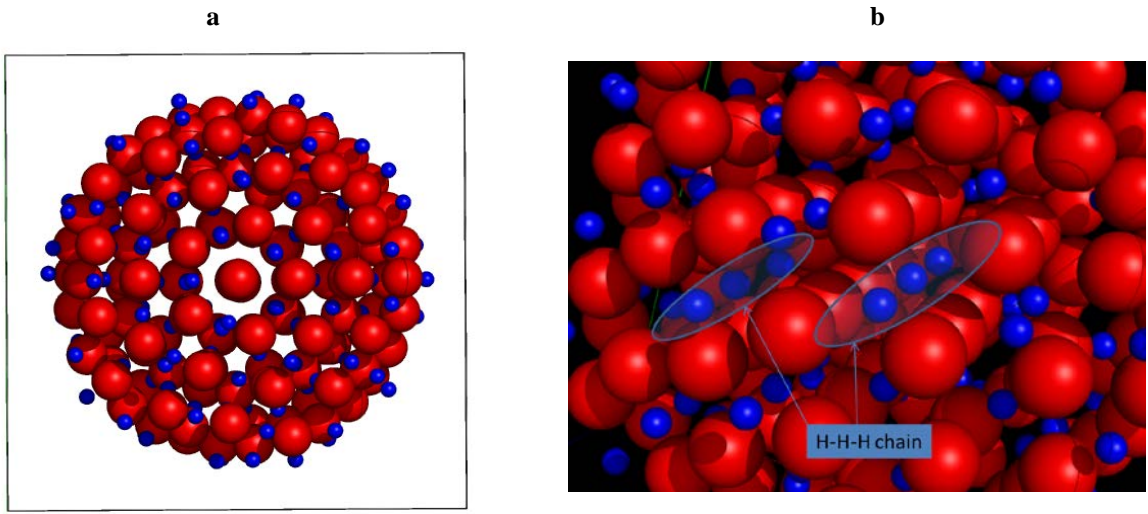

**Fig. 8.** (**a**) Structure of PdH cluster containing 147 Pd and 138 H atoms having minimum free energy configuration, replicated in [33] using the method and parameters by Calvo et al [32]; (**b**) H-H-H chains in the nanocluster, which are viable sites for LAV excitation [34].

## 4. Ways of LAV-LENR triggering

The main message of this paper is that the excitation of LAVs present an efficient way to produce CCS due to time-periodic modulation of the potential well width (or the Coulomb barrier width) and hence to trigger LENR in solids. As has been mentioned above, heating helps to excite LAVs but it decreases their tunneling efficiency. Therefore, we need ways to excite LAVs at sufficiently low temperatures, which is not a trivial problem technically.

One of the methods is based on time-periodic shaking of the surface atoms at frequencies near the optic phonon edge resulting in the LAV excitation in the sub-surface layers. Medvedev et al [35] has demonstrated by means of MD simulations that gap DBs can be excited in the Al sub-lattice of $Pt_3Al$ under the action of *time-periodic external driving*. Time-periodic shaking of the surface atoms at frequencies near the optic phonon edge resulted in the DB excitation in the sub-surface layers. These findings point out at the possibility of LENR stimulation by external time-periodic excitation of surface atoms. This method has been actually realized in "Terahertz" laser experiments [36] on the stimulation of LENR by a joint action of two low-power laser beams with variable beat frequency ranging from 3 to 24THz on the cathode surface during the $D_2O$ electrolysis in the PdD system.

Fig. 9 shows the experimental frequency dependencies of the excess power in these experiments. Three main resonances of excess energy released at $\sim 8\pm1$ THz, $15\pm1$ THz and $21\pm1$ THz have been shown in [8] to correlate with the DB-induced harmonic frequency, $\omega_0(A_{cr}) \approx 7.5$ THz, DB parametric frequency $\Omega(A_{cr}) \approx 15$ THz and DB initial frequency, 21 THz, respectively. According to the model [8], the highest resonance is the biggest, since it is caused by amplification of DB excitation at the edge of optic phonon band. The medium resonance is due to tuning action of external driving on the DB frequencies: it increases the fraction of DBs with parametric frequency. The lowest resonance is due to tuning of harmonic frequencies by external driving: it increases the fraction of D atoms subjected to the parametric action by DBs.

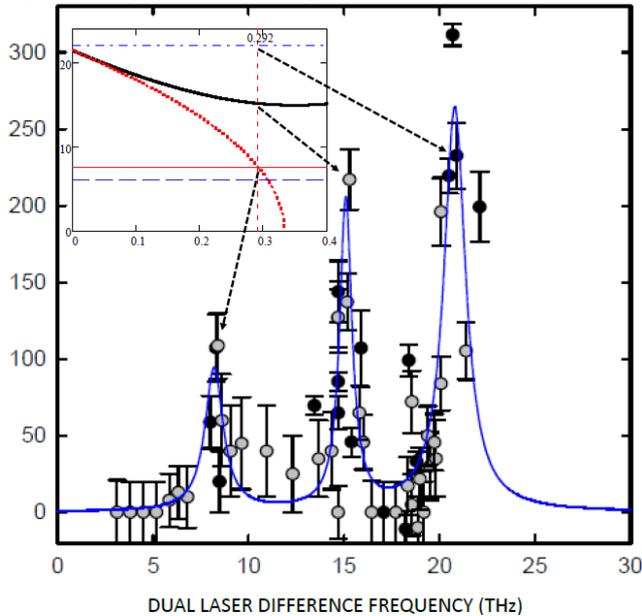

**Fig. 10.** Excess power (mW) under joint action of two low-power laser beams with variable beat frequency on the surface of the Pd cathode during the electrolysis in heavy water [36]. The inset (from the ref [8]) shows parity between critical DB-induced frequencies and the resonance frequencies [36] designated by dashed arrows.

The atoms are shaken by laser beams via excitation of the surface electron plasmons as suggested in [36]. It explains the necessity of external magnetic field for producing resonance effects. However, the direct modulation of the frequency of the optical phonon modes by plasmons proposed in [11] as a driving force for the CCS formation is very questionable [8], and it does not explain LENR *observed in the absence of the laser driving* at slightly higher electric current or temperature [36]. In the present view, the laser driving acts just as a tuning tool for the CCS formation by DBs induced in this case by a joint action of temperature and electric current. Therefore, in order to stimulate LENR in E-cat type installations, one needs to apply some external triggering of LAVs similar to that provided by electric current under electrolysis conditions (below 100 C). Some phenomenological attempts of such triggering are currently under way [37, 38], but the frequency range used in such attempts is far away from the LAV frequencies lying typically in THz diapason. These experiments didn't show any evidence of excess heat within the accuracy of measurements. In the present view, this frequency mismatch may be responsible for the inefficiency of such triggering, which shows the need for *physically based* frequency range and other methods of the **LAV-LENR triggering**.

From the experiments [36] on the resonant enhancement of extra heat generation by dual laser beams, one may conclude that the required frequency range lies in the **terahertz diapason** (8 to 20 THz in the Pd-D system). However, the exact value of the frequency required to trigger LAV-LENR in the Ni-H system *is not known* at present, and if the irradiation frequency does not match the LAV frequency, no effect is expected, in contrast to simple heating that excites all phonon modes. Therefore, the first practical suggestion of the LAV-concept is that one needs a laser with **variable frequency** in the THz range in order to fit the required frequency *in situ* experiments and trigger LENR.

The second suggestion is that instead of the **heating**, which decreases the LAV efficiency, a **cooling** of the nuclear 'fuel' may result in the enhancement of the LENR output. This idea is illustrated in Fig. 11 that shows that the wave lengths corresponding to the desired frequency range (e.g. 6-30 THz) lie in the deep infrared diapason (10-50 microns), which is emitted by a black body having temperature ranging *from the room temperature down to liquid nitrogen temperature* (Fig. 11b).

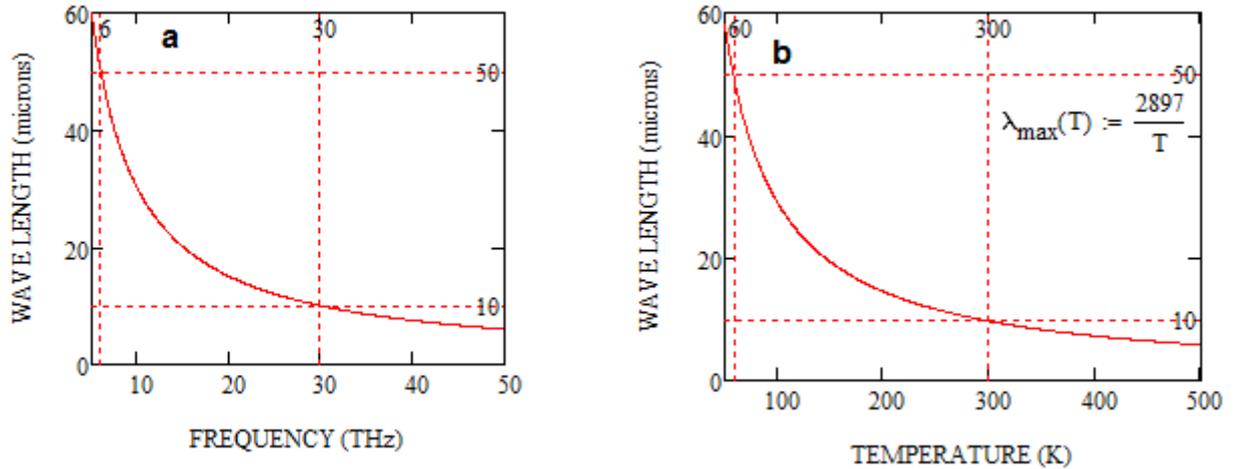

**Fig. 11.** Wave length vs. frequency (**a**) and the maximum wave length vs. temperature of black body (**b**) according to the Wien law (1893).

Thus, a combination of appropriate triggering with cooling may result in the extremely **cold fusion** and other nuclear reactions in a nuclear active environment enriched with potential LAV sites.

## 5. Experimental verification

As has been mentioned above [37, 38], *a majority of attempts to replicate E-cat failed*, which has led J.P. Biberian to the following conclusion [37]: "Reproducing Rossi's experiment following Parkhomov's process is not as easy as it seemed. After more than 20 experiments with nickel and $LiAlH_4$ in different configurations, within the precision of the calorimeter of +/- 2 Watts, no excess heat was measured".

However, there have been a number of successful LENR replications performed by Nick Oseyko [39], who invented a simple method of measuring the effect by comparing the power consumption required to keep the same temperature at the reactor surface with and without 'fuel' ( 0.1 g of $LiAlH_4$). The reaction mixture have been prepared *beforehand* by a slow heating the sealed ceramic tubes containing 1 g of Ni powder and 0.1 g of $LiAlH_4$ a in a muffle furnace up to ~ 800 C.

One of the successful experimental runs performed on 23.07.2015 [40] is shown in Fig. 12(**a**). The PID controller regulated the power consumption from the electrical outlet required to maintain the surface reactor temperature at 1000 C (Fig. 12(**b**)). The control sample (*without fuel*) has shown the mean power consumption of 180 ±20 W (the deviations from the mean value were due to the absence of the voltage stabilization) as shown in Fig. 12(**c**), while the loaded one - 140 ±20 W (Fig. 12(**d**)). This means that the loaded reactor was producing additional heat corresponding to ~ 40W of excess power. No chemical fuel in this amount is capable of such level of heat production, to our knowledge. This particular demonstration lasted 10 min, but according to the inventor (Nick Oseyko), the successful runs have been tested up to 10 hours so far, while the success rate for preloaded samples was about 33% (one out of three samples replicate excess power production at the level of 30 to 40 W). The reason of the failure of the rest 66% of samples to produce the effect remains unknown, being probably the result of different initial microstructure of the fuel.

This controversy of different experimental results points out to the necessity of careful **examination and comparison of microstructures** of successful and unsuccessful fuels, and to the need for search of **new ways of triggering LENR**, which could result in a higher success rate.

One of the perspective ways of triggering LENR is to use *electron beams with **variable beam energy*** [41]. The point is that electrons hitting the target atoms displace them from equilibrium positions by the distance depending on the electron energy and the atomic mass. The displaced atoms start vibrating with frequencies inversely proportional to the initial displacements. If the frequency (i.e. initial displacement) matches the LAV frequency, a LAV is generated. Therefore, electron irradiation with variable energy provides a method of displacing atoms from their equilibrium positions at any desired distance, which in its turn allows one to fit the beam energy to generate LAVs and hence to trigger LENR.

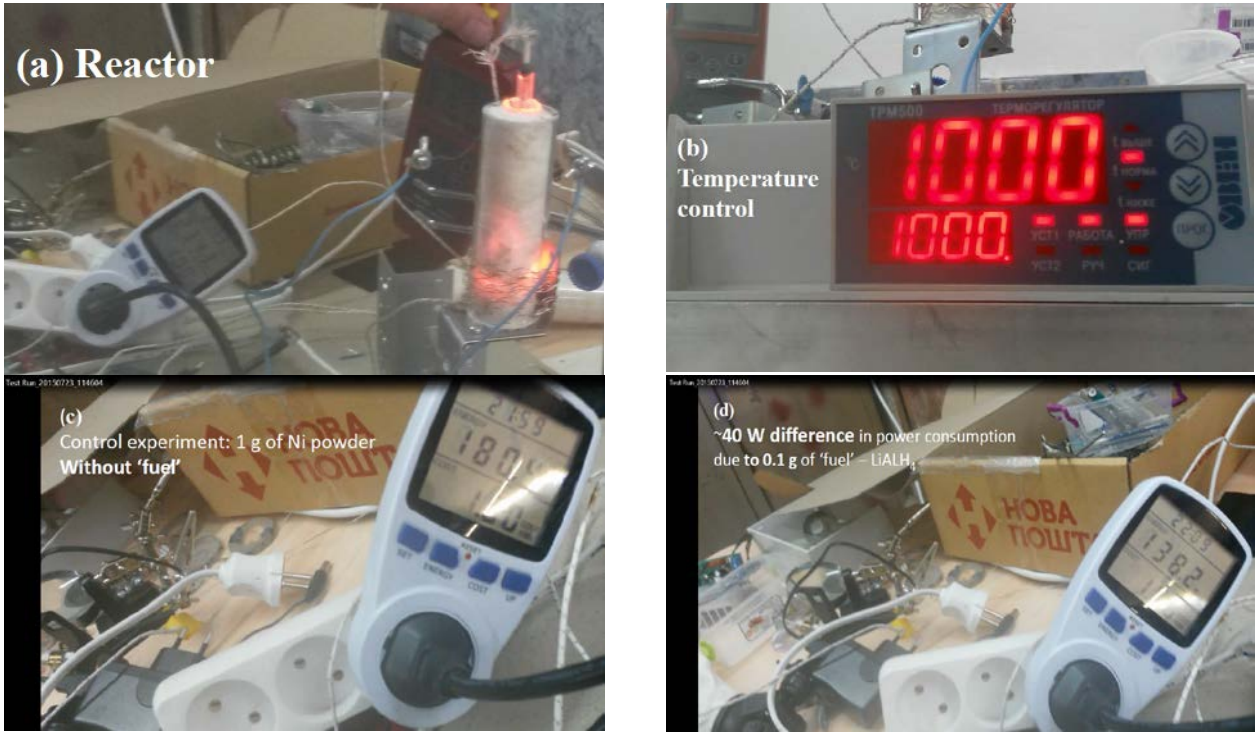

**Fig. 12.** Successful experimental replication of LENR performed by Nick Oseyko [40] (see the text for details).

Fig. 13 shows dependence of the maximum transferred energy of electrons to the target atoms on the beam energy, $E_e$. It can be seen that at $E_e$ ranging from 1 to 100 keV ($E_e$ is the electron energy) the transferred energy ranges from ~ 0.04 eV to 4 eV for Ni atoms and from ~ 2 eV to 240 eV for H atoms. Therefore, it lies well below the threshold energy for permanent damage production (Frenkel pairs) in the Ni case, and it can be higher than that for the hydrogen case. It means that one needs to have an electron source in the range 1-100 keV in order to check all the displacements of interest, which presents a challenging technical problem addressed in the EAVE project [36] and illustrated in Fig. 13. An important advantage of the electron-induced LAV triggering is that the electrons with such energies *penetrate deeply* through the crystal up to the depth ranging from microns to millimeters (as compared to the nm range of laser beam penetration**), which may greatly increase efficiency of the LENR triggering.**

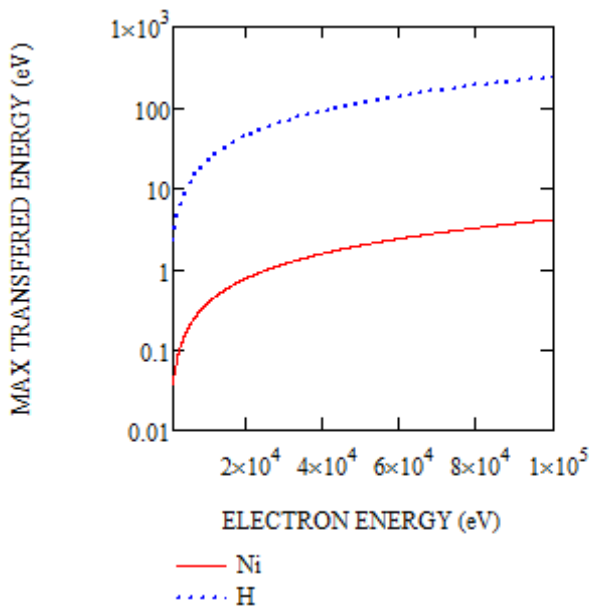

**Fig 13**. Maximum transferred energy of electrons to the target atoms (Ni and H) vs. the beam energy given by

$$E_{\max} = \frac{2E_e\left(E_e + 2m_e c^2\right)}{Mc^2}$$

$E_e$ is the electron energy,

$m_e$ is the electron mass,

$M$ is the target atom mass,

$c$ is the speed of light

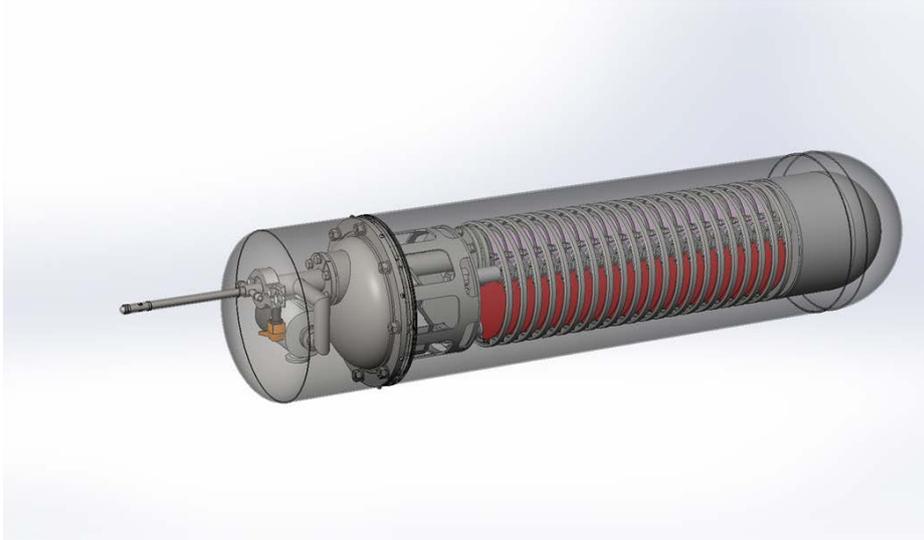

**Fig. 14.** Electron accelerator with variable energy (EAVE) design [41]. The maximum electron current is ~ 1 mA. The dimensions are 94 cm length, 23 cm diameter. The energy consumption is ~ 800 W. The electron energy dispersion in the continuous regime is ~ 0.1%.

## 6. Conclusions and outlook

In the present paper we argue that the excitation of LAVs presents an efficient way to produce CCS due to time-periodic modulation of the potential well width (or the Coulomb barrier width) and hence to trigger LENR in solids. This **LAV concept** is a natural extension of the breather nano-collider (BNC) concept proposed in [8]. In its initial form (Dubinko 2014), it did not take into account quantum correlation effects, and hence, unrealistically small separation between atoms (~ 0.01 Å) would have to be attained in order to enhance the LENR rate up to a noticeable level. Account of the correlation effects (Dubinko 2015) has shown that the *oscillation amplitude of several fractions of angstrom* is sufficient to produce required effect, if CCS parametric conditions are met. Recent results obtained at the US Naval Sea Systems Command [28] verified the present theory by obtaining a direct numerical solution of the time-dependent Schrodinger equation for a single nuclear particle in a parabolic energy well. It has been confirmed that these oscillations in spatial spread of the well will periodically delocalize the nucleus and facilitate the tunneling of adjacent nuclei into the Strong Force attractive nuclear potential well, giving rise to nuclear fusion at rates that are several tens of orders of magnitude larger than what one calculates via the usual Gamow Factor integral relationship.

One of the important practical considerations following from the breather-induced correlation effects is that the **breather lifetime** plays more important role than the minimum distance between the oscillating nuclei.

Another critical parameter is the mean concentration of the CCS produced by the oscillations. In the present paper we extended the concept of **DB**s arising in perfect crystals homogeneously, to the **LAV**s that showed a striking *site selectiveness* of energy localization in the presence of spatial disorder, which has been demonstrated by means of atomistic simulations in biopolymers [5], metal nanoparticles [29] and, *experimentally*, in some *quasicrystals* [9]. The experimentally measured concentration of LAVs in the latter amounts to ~ $10^{20}$ per $cm^3$ that is many orders of magnitude higher than one could expect to find in periodic crystals [8].

Analysis of the LAV frequency range suggests that instead of the currently explored **heating**, which decreases the LAV tunneling efficiency, a **cooling** of the nuclear 'fuel' may result in the enhancement of the LENR output, provided that it is combined with an appropriate **triggering** of the LAV formation by a special **electro-magnetic** or **electron** irradiation.

In conclusion, we note that the present concept is based only on the ***known physical principles*** and on independent atomistic simulations of LAVs in crystals using realistic many-body interatomic potentials and on the *ab initio* structure simulations of Me-H clusters. Atomistic modeling of LAVs of various types in metal hydrides/deuterides is an important outstanding problem since it may offer the ways of *engineering* the nuclear active environment and discover the frequency/amplitude range of LAVs required for the LENR triggering.

**Acknowledgements**

The author would like to thank Denis Laptev for designing Figs. 1, 2 and 6, Sergey Dmitriev – Fig. 4, Dmitry Terentyev – Fig. 8 and Dan Woolridge – LAV animation [34], Klee Irvin – attracting the author's attention to the new and promising field of quasicrystal research. Financial support from Quantum Gravity Research is gratefully acknowledged.